# Specimen preparation for atom probe tomography analysis of complex multifunctional nanoparticles and nanostructures: state-of-the-art and challenges

Varatharaja Nallathambi[1,2*], Se-Ho Kim[2,3], Nikita Polin[2], Natalia F. Shkodich[4], Sven Reichenberger[1], Stephan Barcikowski[1], Baptiste Gault[2,5*]

[1] Technical Chemistry I and Center for Nanointegration Duisburg-Essen (CENIDE), University of Duisburg-Essen, 45141 Essen, Germany

[2] Max Planck Institute for Sustainable Materials, Max-Planck-Str.1, 40237 Düsseldorf, Germany

[3] Department of Materials Science and Engineering, Korea University, Seoul 02841, Republic of Korea

[4] Faculty of Physics and Center for Nanointegration Duisburg-Essen (CENIDE), University of Duisburg-Essen, Lotharstr. 1, 47057 Duisburg, Germany

[5] Univ Rouen Normandie, CNRS, INSA Rouen Normandie, Groupe de Physique des Matériaux, UMR 6634, F-76000 Rouen, France

* corresponding authors: v.nallathambi@mpi-susmat.de, baptiste.gault1@univ-rouen.fr

**Declaration of competing interest**

The authors declare that they have no known competing financial interests or personal relationships that could have appeared to influence the work reported in this paper.

**Abstract**

Atom probe tomography (APT) provides the three-dimensional composition of materials at near-atomic length scales, achieving detection limits in the range of tens of atomic parts-per-million regardless of element type. APT requires the specimen to be shaped as a needle with a tip radius of ~100 nm. The development of site-specific lift-out procedures using focused ion beam–scanning electron microscopy (FIB-SEM) has enabled APT analysis of multifunctional materials, advancing our understanding of their structure-composition-property relationships. Yet these approaches are not readily suitable for analyzing many nanomaterials. Co-electrodeposition of metallic films forms a composite containing the nanomaterials of interest, thereby facilitating APT specimen preparation and enabling analysis of nanowires, nanosheets, and nano- and microparticles, etc. In this perspective article, we showcase diverse examples from simple elementary to compositionally complex alloys of varying dimensionalities, from individual nanoparticles to aerogel structures. We emphasize the challenges encountered with specific material classes during co-electrodeposition procedures and provide recommendations for improving specimen preparation protocols to enhance measurement yield, thereby advancing APT analysis capabilities for optimizing the performance of functional nanomaterials.

**Introduction**

Atom probe tomography (APT) is a time-of-flight mass spectrometer that provides three-dimensional (3D) compositional information with sub-nanometer spatial resolution [1–4]. APT specimens must be sharp and needle-shaped, with typical tip radii below 100 nm to facilitate the field evaporation of surface atoms in the form of ions, thereby enabling their analysis. APT specimen preparation is now typically performed using focused ion-beam scanning electron microscopy (FIB-SEM) [5–7], whereas electrochemical polishing was previously widely employed [8]. The development of site-specific lift-out procedures in FIB has expanded APT's analytical capabilities across a wide range of application areas [5,9,10]. This advancement has enabled the targeting of specific microstructural features such as grain boundaries, phase boundaries, and defect structures for APT analysis within a broad range of bulk material systems [11–14].

Distinctive size and shape effects of nanomaterials provide them with attractive properties for applications spanning biomedicine, electronics, energy storage, and catalysis [15–21]. Since surface-specific structure and composition greatly influence the resulting properties, understanding local structure and chemistry is crucial for developing high-performance nanomaterials for targeted applications. Low-dimensional free-standing nanomaterials presented challenges for the preparation of needle-shaped specimens. However, their significance in functional applications has motivated the development of specimen-preparation workflows that enable APT characterization capabilities. FIB-based specimen preparation for APT analysis of nanomaterials, including nanoparticles and nanostructures with different aspect ratios, requires specialized approaches compared to site-specific lift-out procedures [2]. The most commonly employed method involves embedding free-standing nanostructures into a matrix to facilitate lift-out [22]. Among various approaches explored in multiple studies [22–48], the co-electrodeposition technique for embedding nanomaterials into metallic films has proven particularly effective.

The co-electrodeposition uses a substrate film, typically Cu with a relatively high standard electrode potential, and an aqueous electrolyte of the metal to be deposited (such as Ni, Co, etc.). The nanostructures are either drop-cast/electrophoretically deposited onto the substrate initially or dispersed within the electrolyte during the co-electrodeposition [2,22]. As the metallic film grows on the substrate, it simultaneously embeds the nanostructures. Regions of interest that contain embedded nanoparticles or nanostructures often appear as surface protrusions on the deposited film. Ion milling is subsequently performed to expose the film cross-section and identify regions where nanostructures are embedded. The standard lift-out procedure is then executed to attach a section of the lamella to a silicon post, followed by annular ion milling to prepare needle-shaped specimens. Milling is controlled so that the embedded nanostructure lies close to the tip apex, ensuring inclusions within the analysis volume [22].

In this perspective article, we focus on examples demonstrating how our research group has utilized the co-electrodeposition approach to prepare APT specimens of nanomaterials ranging from simple elementary (Si, AuPd) to compositionally complex nanoalloys (high-entropy nanoalloys) designed for multifunctional applications. We highlight the capabilities of this sophisticated approach while identifying limitations associated with specific material systems. Furthermore, with recent advances in specimen preparation methodologies using *in situ* sputtering developed within our group [49–51], we propose new workflows that could enhance the analysis of free-standing nanoparticles and nanostructures. Complementary methodologies that could strengthen APT-derived compositional information with additional structural data are briefly examined.

Overall, insights gained from analyzing different nanomaterial systems using the co-electrodeposition method have enhanced our understanding of nanomaterial characterization capabilities using APT and defined material-specific limitations. Continued improvements in specimen preparation workflows will advance APT applications for studying and characterizing diverse functional nanomaterials developed for sustainable applications.

**APT analyses of simple to complex nanomaterials**

*Si-based catalyst support*

Support materials for catalysts are crucial for maintaining performance and stability under reaction conditions. Optimized design and fabrication of supports can enhance catalyst performance. An innovative approach to produce conductive catalyst supports for electrode manufacturing in proton exchange membrane water electrolysis (PEMWE) has recently been reported [52]. B-doped Si nanoparticles (p-type doping with a nominal doping concentration of 6 at.% B), ranging from 100-600 nm in size (**Figure 1a**), were synthesized using a hot-wall reactor as support material for Ir catalysts (see Ref. [52] for details).

Here, APT was used specifically to evaluate the distribution of B and other impurity elements within the particles. FIB specimen preparation steps are shown in **Figure S1**. **Figure 1b** shows the 3D atom maps of a B-doped Si nanoparticle embedded in a Ni matrix with an inhomogeneous B distribution. The B content varies in the range of 0.05–0.3 at.% within the particle, and a 1D concentration profile (**Figure 1c**) highlights two segregated planar features indicative of facetted grain boundaries [53–55]. **Figure 1d** shows a close-up of the top region with the respective 1D composition profile (15 nm-diameter cylinder) of elements. A relatively sharp interface (~6 nm) between the Si nanoparticle and the Ni matrix can be observed, which indicates the absence of any elemental dissolution during the co-electrodeposition process. However, the observed 6 nm interface width may be influenced by analytical artifacts, including peak overlap contributions, local magnification effects, and trajectory aberrations [56]. The oxygen is present at the interface, consistent with a thin surface oxide on the nanoparticles that may have prevented dissolution during co-electrodeposition. The relatively large size distribution (100-600 nm) of the Si nanoparticles enabled the easy identification of regions of interest during FIB specimen preparation (**Figure S1**).

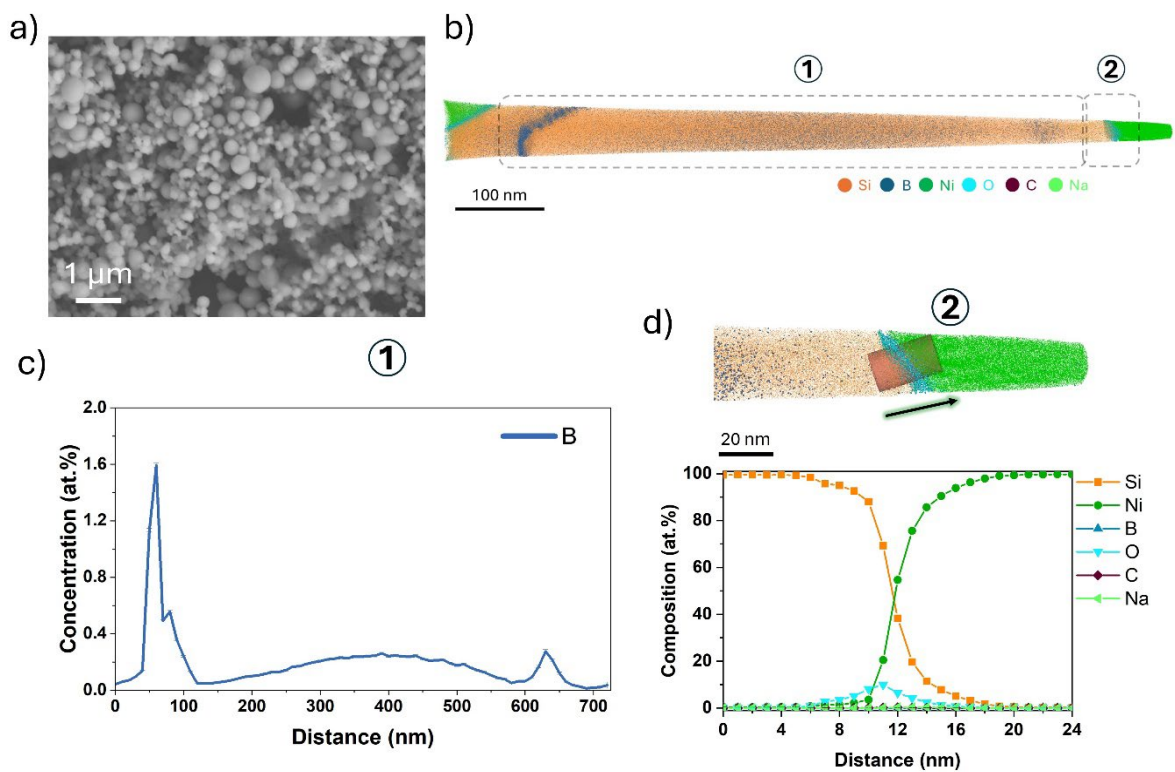

**Figure 1** APT analysis of B-doped Si nanoparticles. a) Secondary electron - SEM micrograph of the B-doped Si nanoparticles; b) APT 3D atom reconstruction of B-doped Si nanoparticle embedded in Ni matrix; c) 1D compositional distribution profile of B across the nanoparticle highlighted as region (1) in (b); d) 1D compositional distribution of elements at the Si-Ni interface highlighted as region (2) in (b) showing the surface oxidized layer. Subfigures (b) and (c) are adapted from Ref. [52], CC BY 4.0 (https://creativecommons.org/licenses/by/4.0/).

*Aerogels for catalysis*

Nanocrystal aerogels are widely explored in catalysis due to their unique characteristics as self-supporting, porous network structures with high surface areas that exhibit good stability under reaction conditions [57–59]. Doping of such aerogel structures results in interesting catalytic behavior and this approach has been used extensively for understanding and tailoring the catalytic performance of monometallic aerogels [43,45,48]. Extending to that, bimetallic aerogels are particularly interesting because of their distinctive catalytic behavior compared to monometallic counterparts [60] and their ability to accommodate various structural configurations, including alloyed, core-shell, and segregated structures [17,47,61,62].

Herein, we focus on our recent study that manipulated three different structural configurations of the Au-Pt bimetallic system through aerogel formation for the glucose cascade reaction [63]. Among the three configurations (segregated, alloyed and core-shell), the segregated Au-Pt aerogels showed enhanced catalytic activity (2.80 and 3.35 times higher than that of the alloy and core-shell variants, respectively). To reveal and validate the structural and compositional arrangement of the bimetallic aerogel system that exhibits optimal activity, APT was employed. Specimen preparation was carried out using the co-electrodeposition method in a Ni matrix (**Figure S2**). Scanning transmission electron microscopy (STEM) characterization results evidencing the segregated structure of the aerogel are shown in **Figures 2a–c** (details in Ref. [63]). A 3D atom map of the aerogel is shown in **Figure 2d**, and a region-of-interest in **Figure 2e,** isosurfaces at 20 at.% Au (blue) and 40 at.% Pt (red) reveals Au-rich regions linked by a Pt-rich network forming a segregated aerogel.

A cross-sectional view and a corresponding 1D concentration plot from a cylindrical region of interest placed across Au–Pt interfaces are shown in **Figures 2f and g**. For the 1D concentration profile, Ni from the matrix was excluded from the calculation. A relatively sharp interface between Au and Pt regions can be observed, indicative of the segregated aerogel structure. A small contribution of Pt in the Au regions could arise due to trajectory aberrations [56,64], which is further evident from the changes observed in the point density (i.e., number of ions per bin) across the interfaces. Additionally, the differences in the relative evaporation field would result in some degree of intermixing [65].

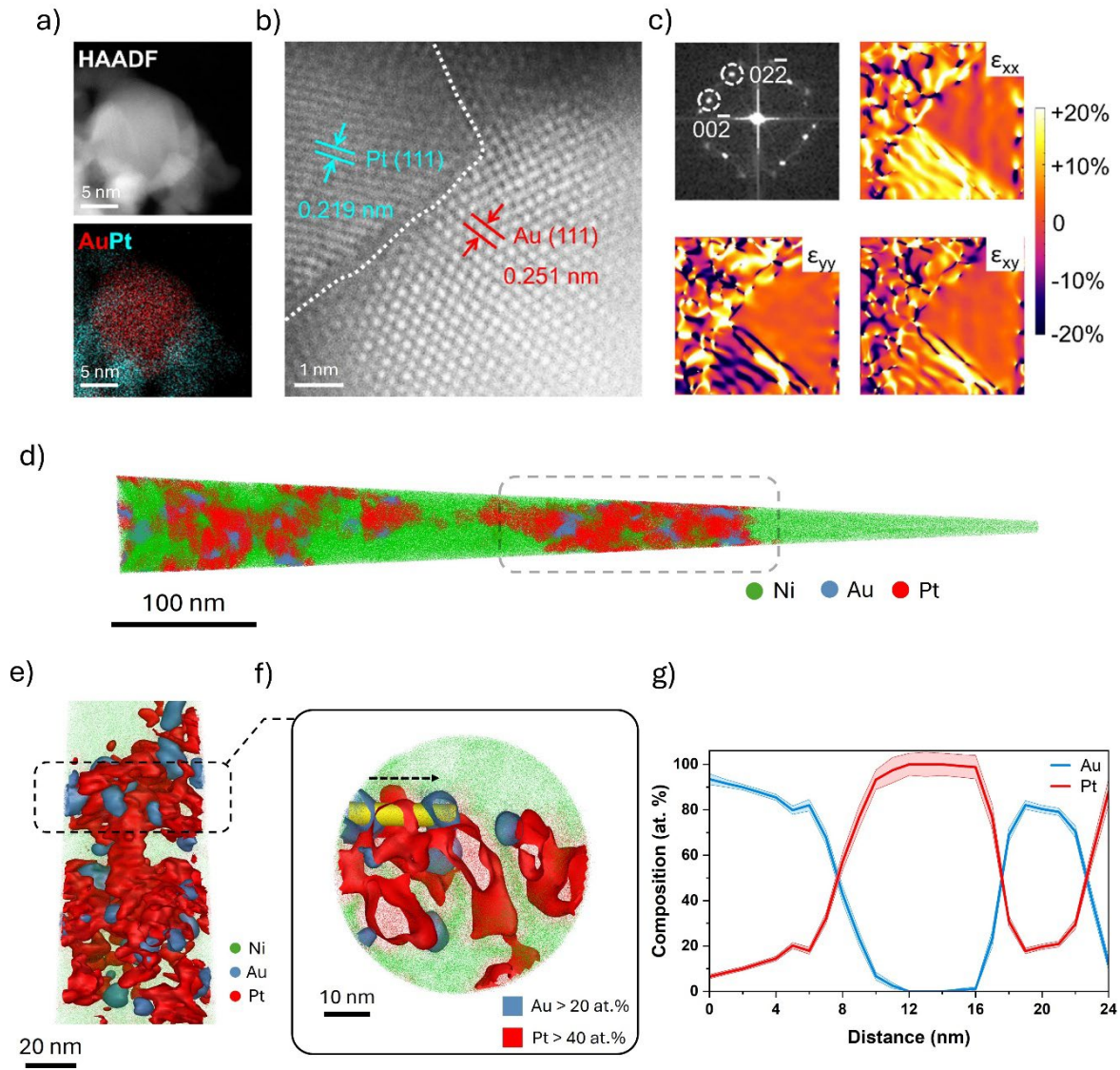

**Figure 2** APT analysis of Au-Pt bimetallic aerogel. a) High-angle annular-dark-field (HAADF) STEM image and the corresponding EDS element distribution maps; b) high-resolution STEM image of the Au–Pt hetero-junction showing the lattice spacings and the Au–Pt phase boundary; c) Fast Fourier Transform (FFT) pattern and the strain distribution maps of (b) along $\epsilon_{xx}$, $\epsilon_{yy}$, and $\epsilon_{xy}$ directions, respectively. A region marked by the white circles in the FFT image is used as the reference during geometrical phase analysis (GPA). The color scale indicates strain changes of −20% to +20%. d) 3D atom reconstruction of the aerogel embedded in a Ni matrix; e) a section of the 3D reconstruction highlighting the segregated structure of the Au-Pt aerogel with isosurface of Au in blue and Pt in red above concentrations of 20 at.% and 40 at.%, respectively; f) cross-section view of the segregated aerogel structure; g) 1D concentration plot using a cylindrical region of interest shown in (f). All subfigures are adapted from Ref. [63], CC BY-NC-ND 4.0 (https://creativecommons.org/licenses/by-nc-nd/4.0/).

*Non-noble metal powder particle*

A notable increase in the coercivity observed for reduction-diffusion processed $ThMn_{12}$-type $(Sm, Zr)_1(Fe, Co, Ti)_{12}$ magnet powders was reported recently [66]. Transmission electron microscopy (TEM) and APT showed monocrystalline powders; a reduced twin-boundary fraction after high-temperature processing (1220 °C) is linked to the enhanced coercivity [67]. APT analyses of the magnet powders of sizes ranging from a few hundred nanometers to 1 μm [66] were facilitated by co-electrodeposition in a Ni matrix [22,44]. **Figure 3a** presents the 3D APT reconstruction of Sm–Zr–Fe–Co–Ti particles processed at 990 °C with the surrounding Ni matrix. A proximity histogram (proxigram) with respect to 50 at.% Ni iso-surface (**Figure 3b**) gives a particle composition $Sm_{5.4}Zr_{1.2}Fe_{71.8}Ti_{6.1}Co_{15.5}$, close to 1:12 stoichiometry [68]. However, the matrix-particle interface extends over approx. 7 nm, with some evidence for an intermixing of the Ni matrix and elements constituting the particle. This suggests dissolution occurring during the co-electrodeposition process and redeposition on the surface of the material, evidencing a limitation of the approach, particularly for small particles from materials susceptible to dissolution.

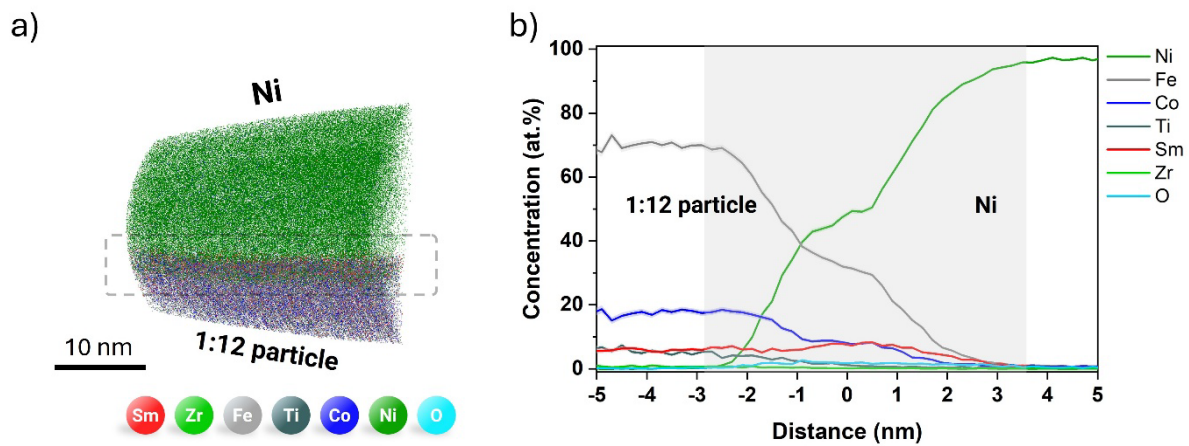

**Figure 3** APT investigation of reduction-diffusion processed $ThMn_{12}$-type Sm–Zr–Fe–Co–Ti magnet powders. a), b) 3D reconstruction and corresponding composition proxigram using a Ni isosurface with a concentration of 50 at.% of the powders processed at 990 °C, respectively. Individual subfigures are adapted from Ref. [67], CC BY 4.0 (https://creativecommons.org/licenses/by/4.0/).

*Transition metal-based HEA nanoparticles*

Laser synthesis and processing of colloids (LSPC) represents a unique approach for producing and tailoring nanoparticles in liquid media, offering advantages including rapid heating and cooling rates, compositional flexibility, and scalability [69–73]. Prior studies have demonstrated HEA nanoparticles with crystallinity spanning amorphous to crystalline, depending on the laser parameters and solvents [74–76]. Solvent selection is therefore critical; our recent study [76] demonstrated pronounced solvent effects on the morphology, structure, and composition of as-synthesized HEA nanoparticles.

For CrMnFeCoNi HEA nanoparticles fabricated by laser fragmentation of ball-milled HEA micropowders (similar to [77,78]), APT results are summarized in **Figure 4**. Electrostatically stabilized colloidals (a major fraction <5nm) agglomerate upon introduction to the electrolyte during Ni co-electrodeposition (**Figure S3**). The 3D atom reconstruction of the HEA nanoparticle assembly is shown in **Figure 4a**. Due to the assembly's larger size, needle-shaped specimens could be prepared without contributions from the Ni matrix. All five HEA constituent elements are present, along with additional oxygen content. Mn-rich regions within the atom map, delineated using a Mn isosurface (>16.5 at.%), are visible in **Figure 4c**. Oxygen enrichment co-localizes with Mn in these regions, indicating preferential surface oxidation during laser synthesis in water [79]. The Mn-O-rich regions appear as a secondary phase within the HEA matrix, while the surrounding areas represent the HEA composition with lower oxygen concentrations. Two isosurface groups were created to distinguish these phase regions (**Figures 4d and e**).

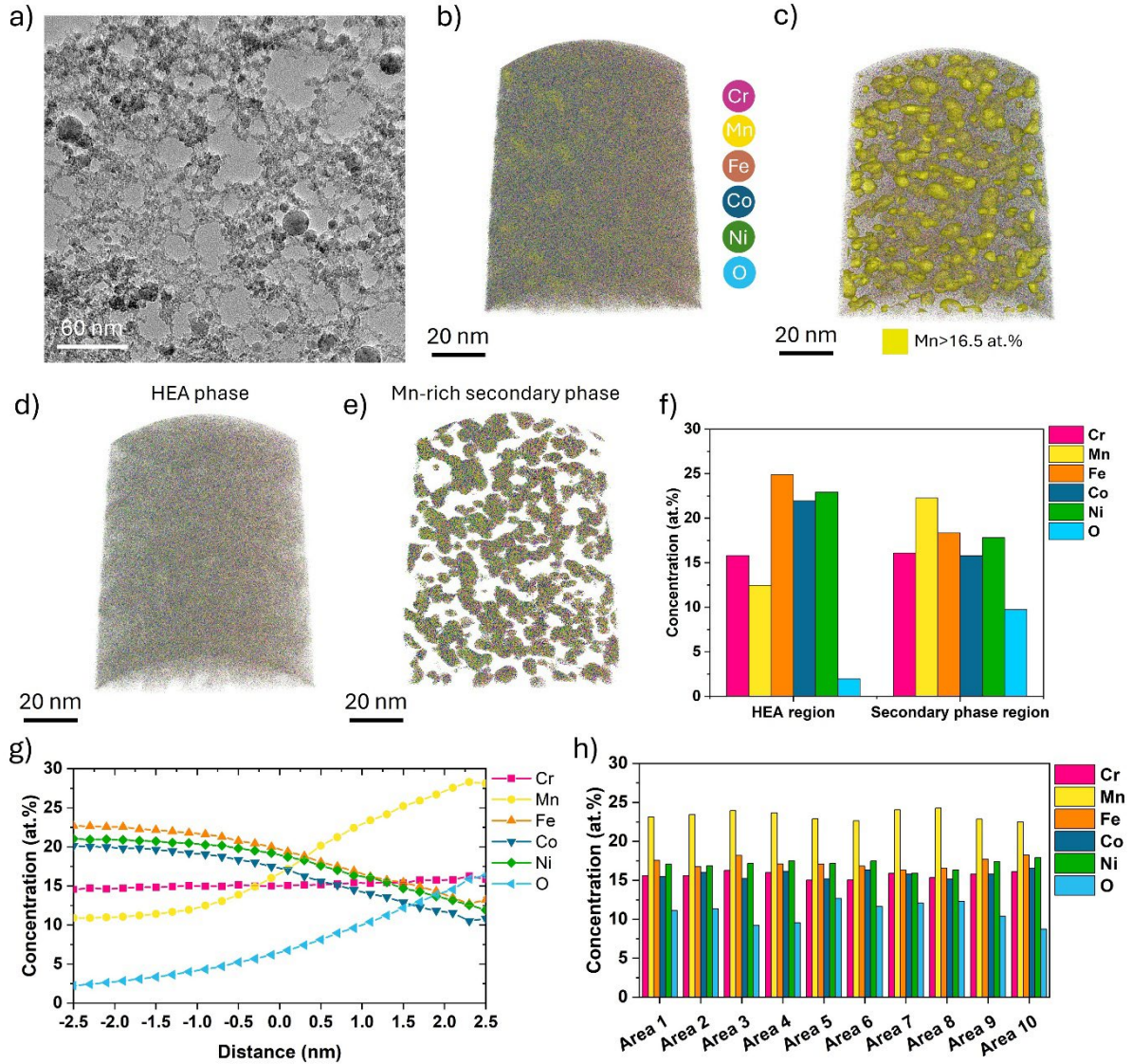

**Figure 4** APT analysis of laser-fragmented CrMnFeCoNi HEA nanoparticles. a) TEM bright field image of the HEA nanoparticles; b) 3D reconstruction of the assembly of nanoparticles embedded in a Ni film; c) Atom map highlighting Mn-rich regions using an isosurface of Mn (yellow) with concentration higher than 16.5 at.%; d), e) Separated atom maps of individual regions corresponding to the HEA and Mn-rich secondary phase, respectively; f) Average composition of elements in the HEA and Mn-rich secondary phase; g) Composition proxigram highlighting the distribution of elements in the Mn-rich secondary phase and the surrounding HEA phase plotted using an isosurface of Mn (>16.5 at.%); h) Average composition of 10 selected Mn-rich secondary phase regions shown in (e).

Average compositions show Mn and O enrichment in the secondary phase, with the remaining elements below equimolar values; the HEA phase is Mn-depleted, with Fe, Co, and Ni slightly above equimolar and minimal O (see **Figure 4e-f)**. Cr appears to be distributed similarly between phases. Proxigram (**Figure 4g**) quantifies partitioning between the Mn-O-rich and HEA phases. The comparable composition of multiple Mn-O-rich regions (**Figure 4h**) suggests an origin in nanoparticle surface oxide layers.

Although as-synthesized nanoparticles likely comprise Mn-rich oxide shells around metallic HEA cores, the 3D reconstruction does not present a representative structure. Instead, Mn-O-rich regions appear as globules embedded within the HEA matrix. This can be attributed to element dissolution and microstructural modifications caused by the acidic nature of the Ni electrolyte (Ni(II) sulfate + boric acid + water) used during co-electrodeposition. Although the larger assembly size facilitates specimen preparation, compositional analysis of individual nanoparticles becomes challenging due to difficulties in delineating interfaces between constituent nanoparticles. In contrast to the passivating oxide layer on the B-doped Si nanoparticles, the thin oxide layers on small nanoparticles of transition metals provide insufficient protection against dissolution.

For HEA nanoparticles synthesized in acetonitrile medium, the nanoparticles are amorphous with a thick carbon shell and minimal surface oxidation (**Figure S4**) [76]. Given that carbon shells could potentially prevent dissolution, a similar co-electrodeposition protocol was employed. Additionally, nanoparticle assemblies were formed prior to the co-electrodeposition process through destabilization during solvent evaporation. The dried nanoparticle assemblies were then embedded in a Ni matrix [22], followed by site-specific liftouts to identify regions of interest within the Ni film for needle-shaped APT specimen preparation. To improve specimen yield, an additional in-situ Cr metallic coating step was incorporated during specimen preparation [49,50]. **Figure 5a** presents the 3D atom reconstruction of the nanoparticle assembly surrounded by in-situ Cr coating. **Figure 5b** delineates the carbon shell region around the nanoparticles using a C isosurface (>1.5 at.%), with nanoparticle regions visualized using a combined Fe-Co-Ni isosurface (>20 at.%). A one-dimensional composition profile generated from a cylindrical region of interest (**Figure 5c**) illustrates the composition distribution between two nanoparticles with an intervening C-rich shell. The carbon concentration within particles ranges from 12–15 at.% with uniform Fe, Co, and Ni distribution, and the thick carbon shells potentially shield nanoparticles from dissolution while improving delineation between individual particles.

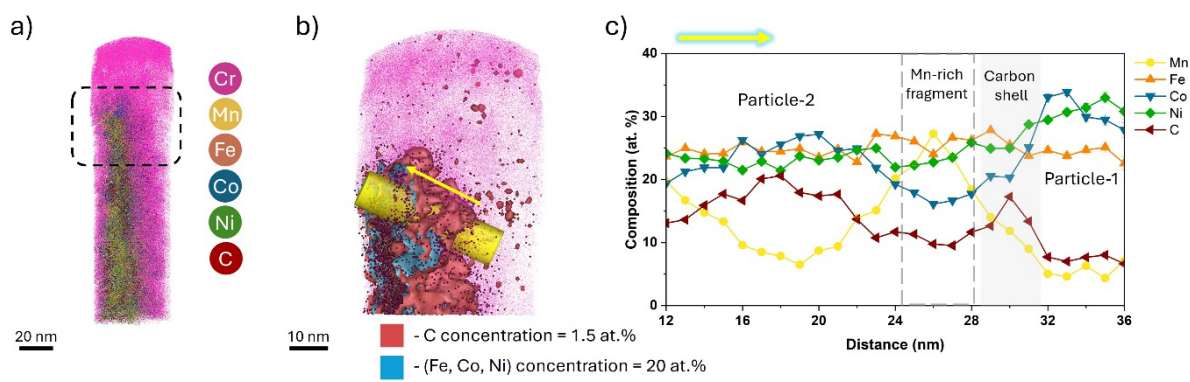

**Figure 5** a) 3D APT reconstruction of laser-fabricated amorphous HEA nanoparticle assembly encapsulated in an in-situ coated Cr layer; b) Atom map from a selected region highlighting the carbon shell surrounding the nanoparticle region using an isosurface (brown) of C with a concentration greater than 1.5 at.%; c) 1D composition profile from the selected region of interest shown in (b) highlighting the composition distribution as a function of distance. Reproduced from Ref. [76], CC BY 4.0 (https://creativecommons.org/licenses/by/4.0/).

Additional efforts were undertaken to enable single-particle analysis by using substrates with rough surface morphology to improve nanoparticle adhesion. The colloidal nanoparticles were drop-cast onto the substrate surface and air-dried, followed by co-electrodeposition. An example (**Figure 6**) shows the analysis of CrMnFeCoNi HEA nanoparticles prepared by laser synthesis in ethanol (similar to [76,80]) and co-electrodeposited on an alumina-coated Cu substrate (**Figure S5**). Co was selected for electrodeposition based on previously promising results [46,81], as it exhibits similar evaporation field characteristics to Ni and produces a single isotope peak in the APT mass spectrum compared to five peaks for Ni. The APT reconstruction illustrating the distribution of nanoparticle regions is shown in **Figures 6a and b**, with and without Co contribution, respectively. Evidence of elemental dissolution is observable in **Figure 6b**, while close-up views of the nanoparticle regions are highlighted in **Figures 6c and d** using a combined isosurface of Cr-Mn-Fe-Ni (> 15 at.%). The relative elemental concentrations of the nanoparticle regions marked in **Figures 6c and d**, excluding Co, are presented in **Figure 6e**. The distribution appears non-uniform with Cr-, Ni-, and Mn-rich particles, indicative of possible dissolution during co-electrodeposition. However, the larger nanoparticle visualized in **Figure 6f** appears intact, and the corresponding 1D concentration profile showing individual element distribution is provided in **Figure 6g**. A sharp particle/matrix interface is observed where a thin Mn-enriched oxide layer separates the Cr-rich nanoparticle from the Co matrix, suggesting possible dissolution prevention by surface oxide layers in larger nanoparticles and enabling individual particle analysis.

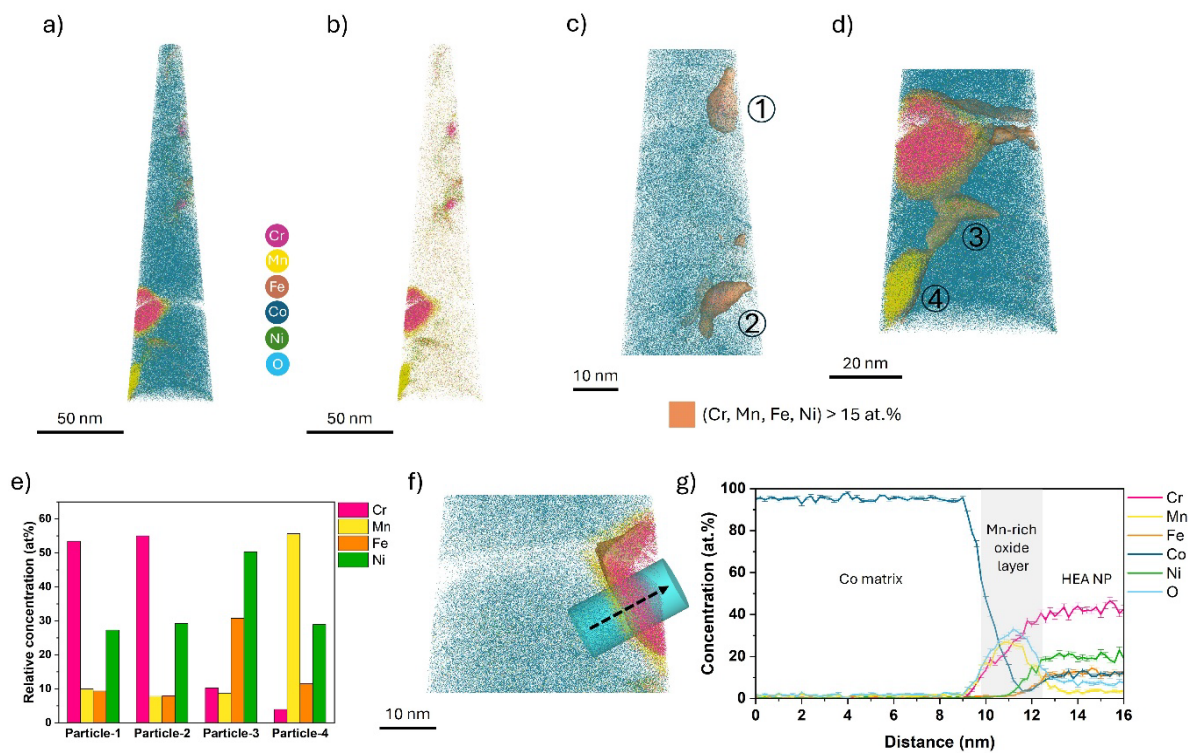

**Figure 6** APT analysis of laser-generated HEA nanoparticles embedded in a Co film. a), b) 3D atom reconstruction showing the regions of nanoparticles with and without Co matrix, respectively; c), d) close-up of the nanoparticle regions; e) relative composition of individual nanoparticles excluding Co, marked in (c) and (d); f), g) 1D composition profile showing the distribution of individual elements in the highlighted nanoparticle.

**Limitations, perspectives and improvements**

While co-electrodeposition using a Ni/Co matrix effectively embeds nanostructures for most systems discussed above, several limitations warrant consideration. Site-specific lift-out procedures using FIB are most successful for microparticles, larger nanoparticles (diameter >200 nm), 2D structures and nanoparticle assemblies such as aerogels, as larger nanostructures provide better contrast for identifying regions of interest within the deposited film. However, individual nanoparticle analysis has gained increasing interest, particularly for catalytic applications, where compositional information of the outermost atomic layers is crucial. This presents challenges with current sample preparation methods, requiring careful measures to prevent agglomeration or aggregation during co-electrodeposition. Additionally, analysis of smaller individual nanoparticles, especially those <20 nm in diameter, is challenging due to identification difficulties within the deposited metallic film during FIB specimen preparation.

Furthermore, dissolution or surface modification may occur during the co-electrodeposition process, particularly with non-noble materials, potentially altering the native nanoparticle's characteristics, including microstructure, composition, shape, etc. Intentional agglomeration or formation of nanoparticle assemblies (e.g., via gelation to form aerogels [63,82]) prior to electrodeposition could potentially minimize dissolution effects and improve their identification in the deposited film. However, this approach may compromise the delineation of individual nanoparticles and interface identification. Conductive support materials such as metal microparticles could address the contrast limitations during FIB preparation. Supported nanoparticle catalysts on conductive supports encapsulated in metallic films can facilitate easy identification of regions of interest and potentially prevent individual nanoparticle agglomeration during co-electrodeposition. Experimental trials by our group to encapsulate laser-fabricated colloidal nanoparticles with a thin $SiO_2$ surface layer using the sol-gel method, inspired by Ref. [83], were attempted but proved unsuccessful, possibly due to specific interactions between nanoparticles in the colloidal solution or their modified surface chemistry after laser synthesis. In the past, atomic layer deposition (ALD) of a 3 nm-thick $Al_2O_3$ layer on ZnS particles was successfully performed in a rotary drum reactor [84], but the transferability of the ALD method towards coating oxides or metals appears hardly straightforward and will potentially require extensive optimization.

Further improvements to the co-electrodeposition process can address dissolution issues through several approaches. First, minimizing the applied potential window during nanoparticle contact with the electrolyte can reduce elemental dissolution. This can be combined with cathodic pre-bias, where an initial metallic film layer is deposited on the substrate before introducing nanoparticles into the electrolyte, enabling more uniform and dense embedding. Pulse electrodeposition protocols have demonstrated improved nanoparticle number density compared to DC electrodeposition [85]. Additionally, transitioning to low-chloride sulfamate-type electrolytes, non-aqueous alternatives, or exploring electroless plating methods [86,87] warrant further investigation. Finally, selecting a matrix material with an evaporation field closely matching that of the nanoparticles is crucial for minimizing local magnification artifacts [56].

Another aspect to consider is measurement yield. In the cases mentioned above regarding individual particle analysis, the yield is considerably lower compared to aerogel nanostructure analysis because of the difficulty in finding the regions of interest during FIB preparation. To address this limitation, electrophoresis or sputter deposition techniques could be employed to uniformly distribute nanoparticles onto the substrate surface, followed by co-electrodeposition for encapsulation [88]. This approach provides a direct substrate-deposited film interface as the region of interest for APT specimen preparation and could potentially increase measurement yield for single nanoparticle analysis. Furthermore, an additional *in situ* metallic coating step (such as Cr or Pd) on prepared APT specimens can not only improve measurement yield but also enhance data quality, increase field of view, and improve mass resolution [49,50], which has also been shown to be beneficial in the case of 2D materials recently [51]. Additionally, the choice of electrodeposited metal can be optimized to minimize overlaps with nanoparticle composition. Ni and Co are commonly selected due to their ease of electrodeposition and close field evaporation, matching most elements of interest. Alternative options include Cu [89,90] and Zn, while Pt, Mo, and Ag could also be considered.

Beyond metallic film co-electrodeposition for nanoparticle embedding, studies have reported electrophoresis of nanoparticles onto pre-sharpened APT needles or direct sputtering of nanostructures or thin films onto pre-sharpened Si tips [30,31,35,39,91,92]. A specimen preparation workflow suitable for nanoparticles in dispersion reported previously [93] yielded reasonable throughput, which involved the electrophoresis on a presharped APT tip followed by Ni electrodeposition. However, analysis from such preparation methods can be challenging, as they are prone to shadowing effects and residual microvoids or imperfections may cause early specimen fracture or lead to reconstruction artifacts. An improved approach involves *in situ* metallic sputtering of substrates in FIB, where nanoparticles are uniformly distributed on the surface. A schematic illustrating the recommended steps involved during the APT specimen preparation using the *in situ* sputtering method for nanoparticles is shown in **Figure 7**. However, careful consideration must be taken to ensure nanoparticles remain intact and adhere to the substrate surface during the coating process without damage or displacement from electron/ion beam exposure. Other possible solutions that are, however, not yet reported could be to electrophoretically deposit the nanoparticles on a pre-sharpened tip, analogously to [39,94,95], when intending to investigate solely electrostatically stabilized colloidal nanoparticle suspensions via APT, followed by a similar in situ sputtering protocol shown in **Figure 7**.

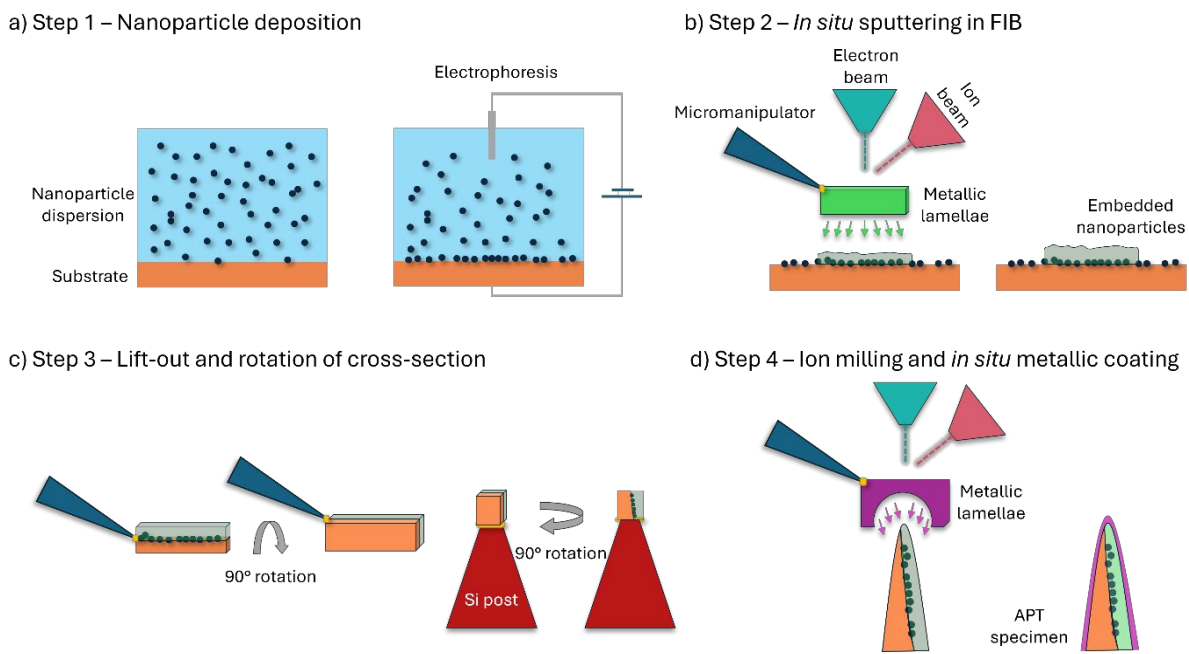

**Figure 7** Schematic illustration of the recommended steps involved during the APT specimen preparation using the *in situ* sputtering method for nanoparticles. a) Deposition of nanoparticles onto a substrate using the electrophoresis techniques; b) *In situ* sputtering of the nanoparticle-deposited substrate surface using a metallic lamella to embed the nanoparticles; c) Site-specific lift-out of the sputtered surface and rotation of the lamella cross-section to align the substrate-sputter coating interface; d) Annular ion milling followed by *in situ* metallic coating of the needle-shaped APT specimens.

**Future prospects and complementary methodologies**

Despite limitations arising from specimen preparation outlined above, APT has enabled compositional analysis of simple to complex nanomaterials at near-atomic length scales, where the quantification of certain impurity elements is otherwise difficult with other analytical techniques [96]. Dopant and impurity concentration quantification has proven effective using APT in functional nanomaterials. Despite the effectiveness of embedding nanomaterials in metallic films, exploring alternative embedding methods provides additional flexibility for material classes prone to surface restructuring or modification during electrodeposition. Recent advances in cryo-based specimen preparation protocols [50,97] supported with *in situ* sputtering [49,51] can enable embedding of sensitive nanomaterials while preserving their surface chemical environment. Additional surface protection steps can be implemented prior to cryo or chemical fixation to prevent surface degradation.

Although certain nanostructure analyses face limitations, the compositional data gained from APT provides immense value. An effective approach involves complementing or coupling

APT with other analytical tools to obtain high-resolution structural and compositional information. Correlative studies combining APT and transmission electron microscopy have gained prominence to facilitate functional nanomaterial analysis, particularly by adding structural information to APT-derived compositional data. Additional techniques such as atomic electron tomography, X-ray photoelectron spectroscopy, secondary ion mass spectrometry and high-resolution X-ray and neutron diffraction show promising potential. The drive toward sustainable energy sources and practices necessitates the design and fabrication of advanced, efficient functional nanomaterials, where near-atomic-scale structural and compositional characterization tools play a crucial role in tailoring functional properties.

**Conclusions**

The development of novel, high-performance functional nanomaterials requires rigorous experimental investigations involving microstructural and spectroscopic characterization tools to correlate the structure-property relationships. Atom probe tomography of nanomaterials advances the development cycle by providing compositional information at near-atomic length scales, with specimen preparation representing the primary bottleneck. We have summarized potential APT applications in analyzing various nanostructures, including compositionally complex alloys containing six elements, that were performed in our group, ranging from nanoparticles to aerogels, using co-electrodeposition. Additionally, material-specific limitations and challenges arising during co-electrodeposition are highlighted, and potential solutions to overcome these setbacks, along with new specimen preparation methods for individual particle analysis, are proposed. Recent developments in site-specific lift-outs and *in situ* sputtering of metallic films within FIB will further advance workflows and improve the efficiency of nanomaterial analysis using APT in the coming years.

**Acknowledgments**

V.N. is grateful for the financial support from the International Max Planck Research School for Interface Controlled Materials for Energy Conversion (IMPRS-SurMat), now International Max Planck Research School for Sustainable Metallurgy (IMPRS-SusMet), and the Center for Nanointegration Duisburg-Essen (CENIDE). The authors are grateful to Uwe Tezins, Christian Broß, and Andreas Sturm for their support at the APT, SEM, and FIB facilities and thank Philipp Watermeyer and Volker Kree for their support at the TEM facilities at the Max Planck Institute for Sustainable Materials. V.N. greatly appreciates Mathias Krämer, Mahander Pratap Singh, Ezgi Hatipoğlu, Aparna Saksena and Eric Woods at the Max Planck Institute for Sustainable Materials for their assistance and helpful discussions on FIB specimen preparation and APT analysis. S.-H.K. acknowledges the support from the National R&D Program through the National Research Foundation of Korea (RS-2025-00520824). N.F.Sh. and B.G. acknowledge the financial support from the Deutsche Forschungsgemeinschaft (DFG) within CRC/TRR270, projects A04, Z01 (Project ID 405553726), and project FA209/27-1.

# Supporting Information

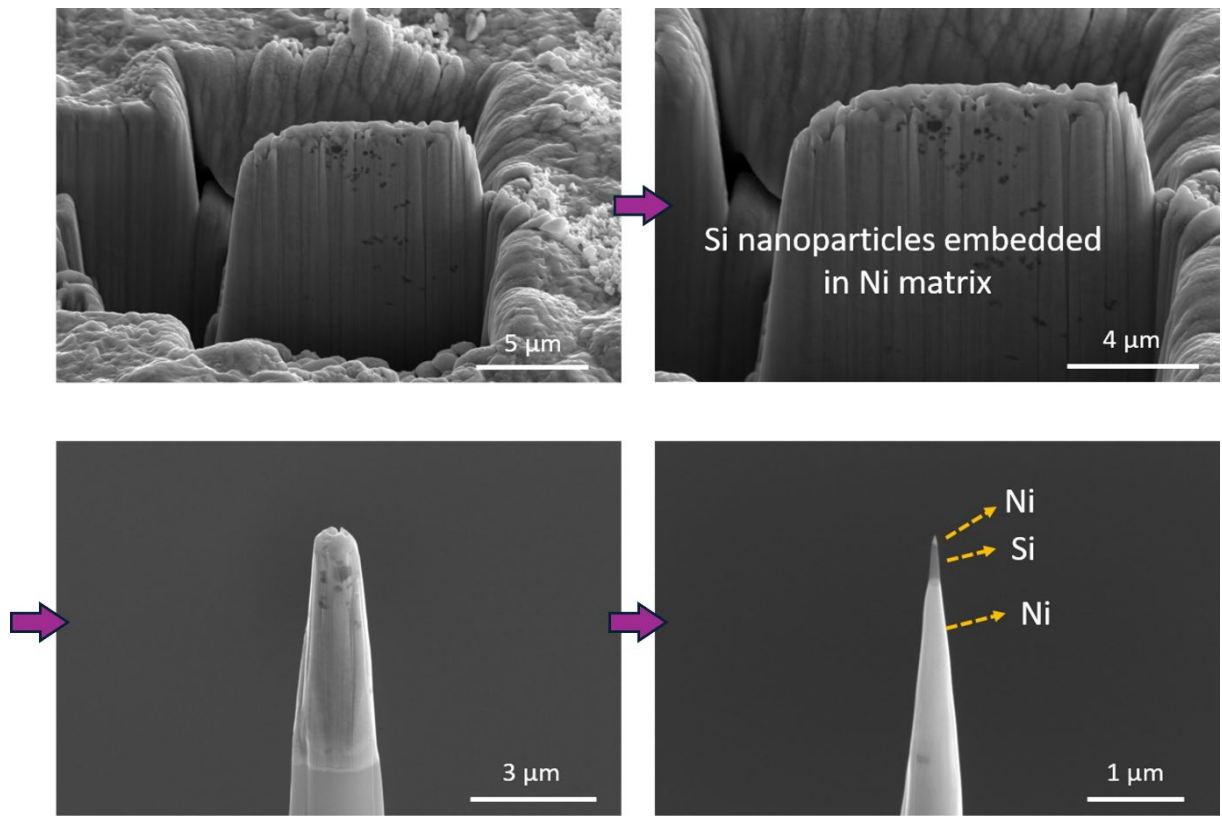

**Figure S8** APT specimen preparation steps for the Si nanoparticles embedded in a Ni film using the standard site-specific lift-out procedure in FIB. Reproduced from Ref. [52], CC BY 4.0 (https://creativecommons.org/licenses/by/4.0/) .

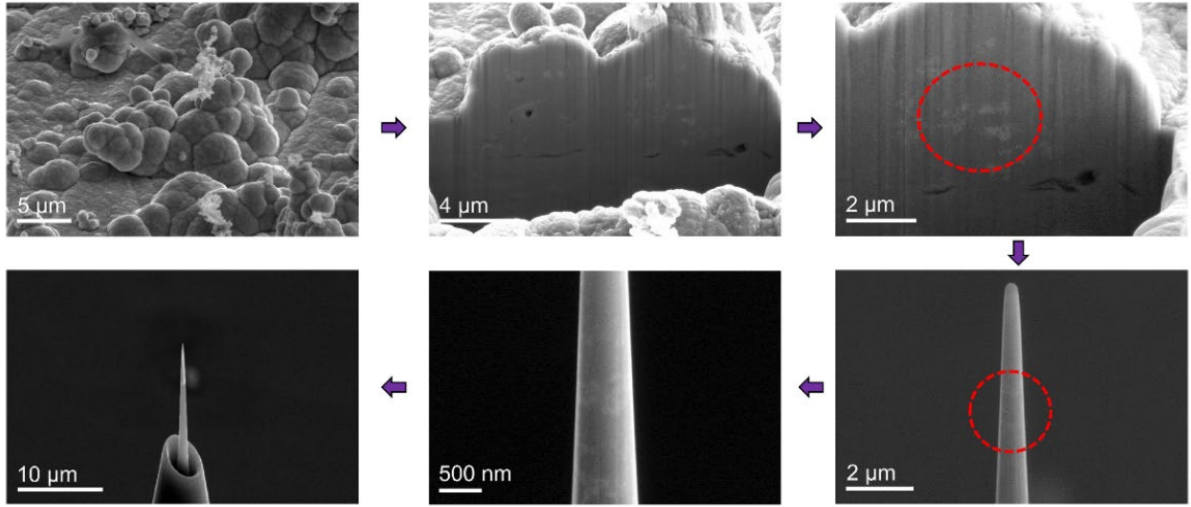

**Figure S9** Steps involved in the APT specimen preparation from a co-electrodeposited Ni film using the standard site-specific lift-out procedure in FIB. The embedded aerogel regions are highlighted in red circles. Reproduced from Ref. [63], CC BY-NC-ND 4.0 (https://creativecommons.org/licenses/by-nc-nd/4.0/).

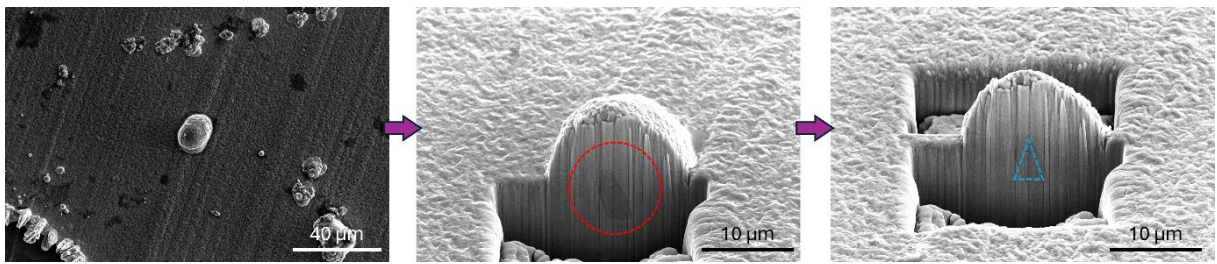

**Figure S10** SEM images showing the agglomerated nanoparticle region (red and blue) identified within the cross-section of the electrodeposited film during the standard site-specific lift-out procedure in FIB.

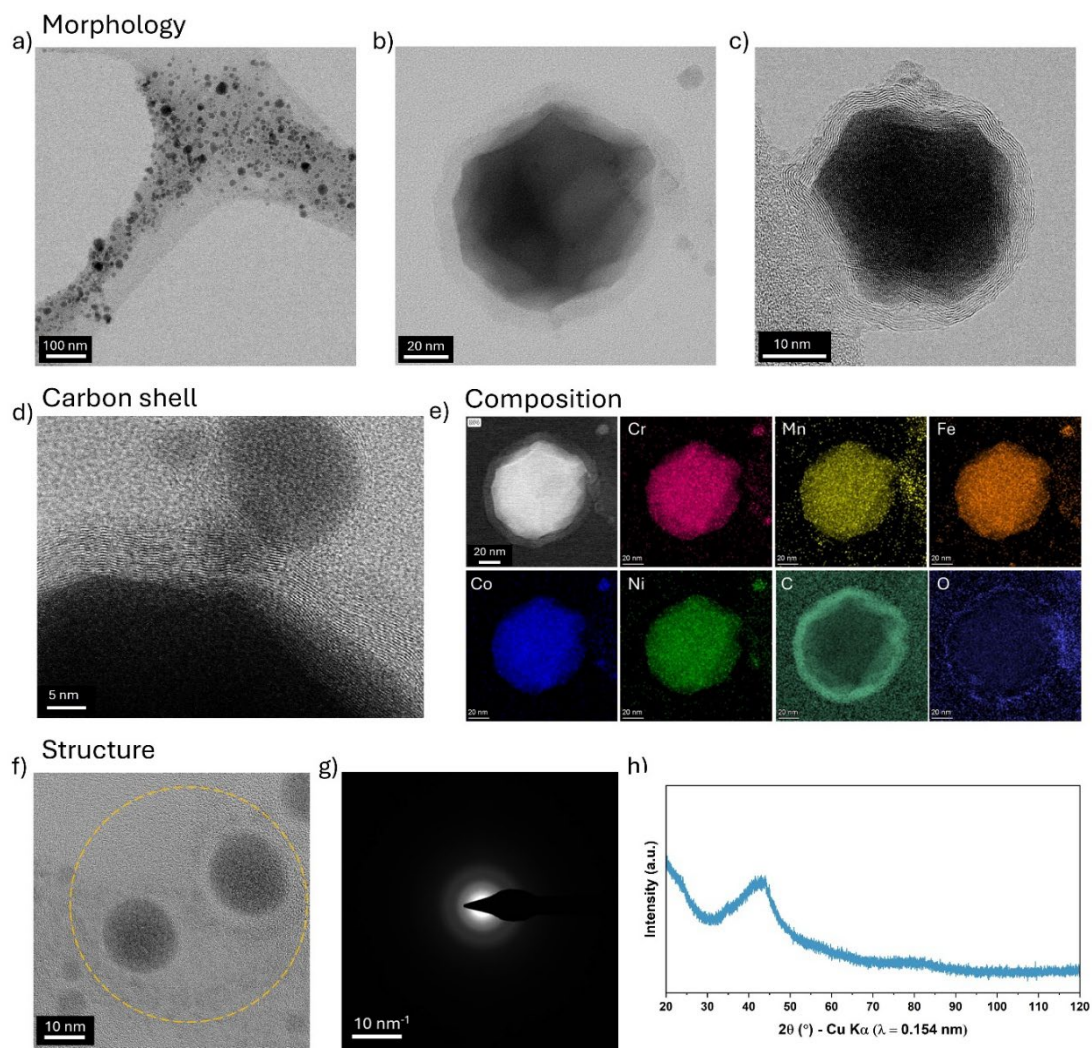

**Figure S11** Morphology, composition, and structural characterization of HEA NPs synthesized in acetonitrile. a) TEM bright field image showing the nanoparticles' morphology and size variations; b), c) STEM bright field images of selected nanoparticles of different sizes highlighting their rugged surface covered with a carbon shell around; d) STEM bright field image at a higher magnification showing the onion-like morphology and size-dependent carbon shell thickness differences between two NPs; e) STEM-EDS analysis showing the individual element distribution maps; f), g) TEM bright field image and its respective SAED pattern highlighting the amorphous nature of the NPs; h) Powder XRD pattern of the NPs showing a diffused peak indicating their amorphous nature. Reproduced from Ref. [76], CC BY 4.0 (https://creativecommons.org/licenses/by/4.0/).

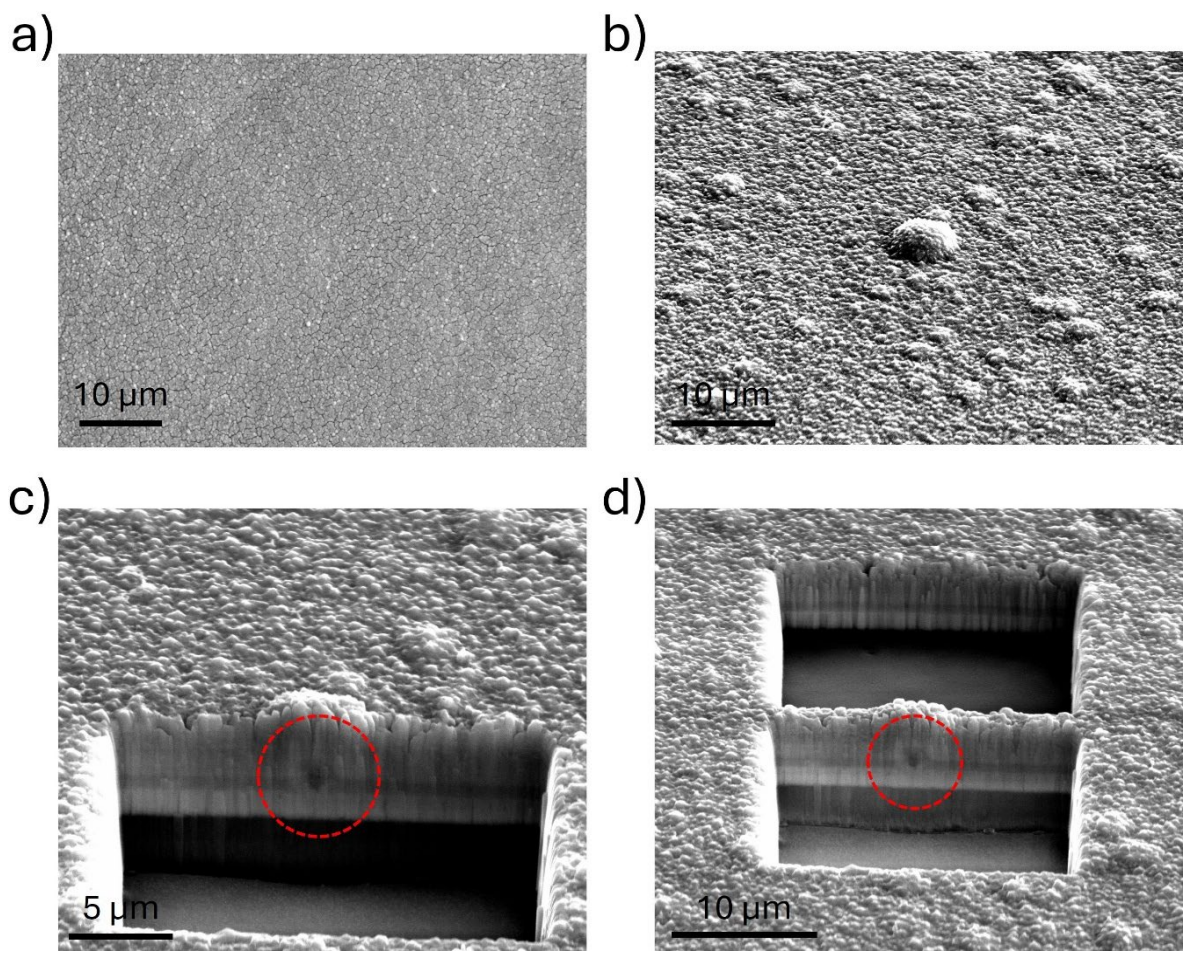

**Figure S12** a) SEM image of the Al$_2$O$_3$-coated Cu substrate with dropcasted nanoparticles; b), c), d) SEM images showing the embedded nanoparticle regions (red) in co-electrodeposited Co film on the Al$_2$O$_3$-coated Cu substrate.